\documentclass[sigconf, screen, anonymous=false,authorversion,nonacm]{acmart}

\newcommand{\adder}[0]{AD{$\Delta$}ER}

\usepackage[capitalize]{cleveref}
\usepackage{caption}
\usepackage{subcaption}
\usepackage{float}
\usepackage[htt]{hyphenat} 
\usepackage{array}
\usepackage{arydshln}

\usepackage{svg}
\usepackage{algorithm}
\usepackage{algpseudocode}
\usepackage{pifont}
\usepackage{lipsum}  
\usepackage{diagbox}
\usepackage{multirow}

\crefname{section}{Sec.}{Secs.}
\Crefname{section}{Section}{Sections}
\Crefname{table}{Table}{Tables}
\crefname{table}{Tab.}{Tabs.}

\usepackage{hyperref}
\hypersetup{
    colorlinks=true,
    linkcolor=blue,
    filecolor=magenta,      
    urlcolor=cyan,
    pdftitle={Overleaf Example},
    pdfpagemode=FullScreen,
    }
\usepackage{graphicx}

\AtBeginDocument{%
  }

\setcopyright{acmcopyright}
\copyrightyear{2024}
\acmYear{2024}
\acmDOI{XXXXXXX.XXXXXXX}

\acmConference[MMSys '24]{Proceedings of the 15th ACM Multimedia Systems Conference}{April 15--18,
  2024}{Bari, Italy}
\acmBooktitle{Proceedings of the 15th ACM Multimedia Systems Conference (MMSys '24), April 15--18,
  2024, Bari, Italy}
\acmPrice{15.00}
\acmISBN{978-1-4503-XXXX-X/18/06}

\begin{document}

\title{An Open Software Suite for Event-Based Video}

\author{Andrew C. Freeman}
\email{acfreeman@cs.unc.edu}
\orcid{0000-0002-7927-8245}
\affiliation{%
  \institution{University of North Carolina}
  \city{Chapel Hill}
  \state{North Carolina}
  \country{USA}
}

\renewcommand{\shortauthors}{Freeman}

\begin{abstract}
  While traditional video representations are organized around discrete image frames, event-based video is a new paradigm that forgoes image frames altogether. Rather, pixel samples are temporally asynchronous and independent of one another. Until now, researchers have lacked a cohesive software framework for exploring the representation, compression, and applications of event-based video. I present the \adder{} software suite to fill this gap. This framework includes utilities for transcoding framed and multimodal event-based video sources to a common representation, rate control mechanisms, lossy compression, application support, and an interactive GUI for transcoding and playback. In this paper, I describe these various software components and their usage.
\end{abstract}


\begin{teaserfigure}
  \centering
  \includegraphics[width=0.56\textwidth]{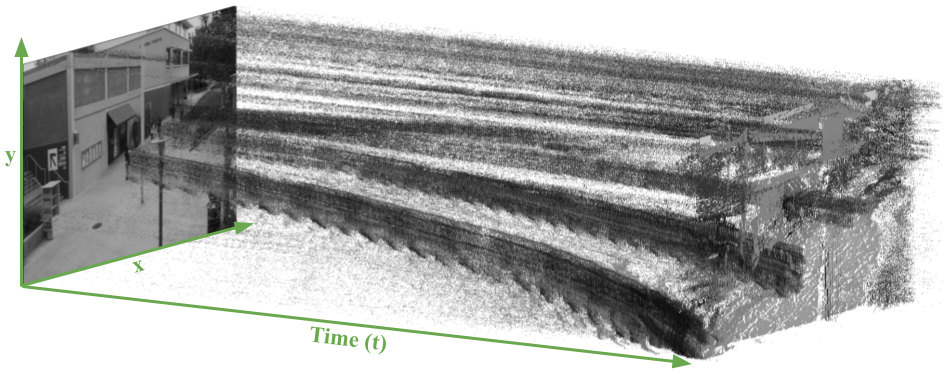}
  \caption{Visualization of an \adder{} video. The framed input (left) produces a stream of temporally sparse intensity samples (``events'') which are concentrated on areas of high motion. The burst of events near the beginning (right) ensures that we obtain an initial intensity for every pixel.}
  \label{fig:teaser}
\end{teaserfigure}

\received{20 February 2007}
\received[revised]{12 March 2009}
\received[accepted]{5 June 2009}

\maketitle

\section{Introduction}

The human eye does not sense the world with image frames. Due to the sensing hardware of traditional cameras, however, classical video representations are fundamentally a sequence of discrete images through time. \textit{Event video}, on the other hand, is a representational paradigm that eschews image frames in favor of independent and asynchronous pixel streams.

To date, most research for event video has focused on event-based sensing hardware and creating custom applications for specific cameras. Often, the practical concerns of compression, rate adaptation, and generic applications are ignored entirely. Frame-based video systems, where these concerns are well known and well addressed, have likewise not seen a thorough exploration of the consequences of event-based representation. I argue that by transcoding certain types of frame-based video (e.g., surveillance and slow motion) to an event representation, one can see enormous improvements to compression performance and vision application speed. Furthermore, event representations are amenable to spiking neural networks (SNNs), a rapidly growing research area in the realm of neuromorphic computing.

With this open-source release, I separate the concerns of researchers in the disparate areas of event-based hardware, networking, and vision: rather than developing techniques for a single event camera and file format, researchers' efforts can have forwards compatibility with future camera types. Furthermore, I bring classical video into the asynchronous paradigm, meaning that event-based applications developed for this framework will also be compatible with traditional frame-based video sources. In this paper, I describe in detail my software for transcoding to a unified event representation, rate adaptation, compression mechanisms, event-based applications,  visual playback, stream inspection, and a graphical interface. The software is available from a centralized repository at \href{https://github.com/ac-freeman/adder-codec-rs}{https://github.com/ac-freeman/adder-codec-rs}.

\section{Related Work}



\subsection{Event Cameras}

In recent years, neuromorphic ``event'' camera sensors have entered the fold for robotics research. Rather than recording image frames, pixels within these sensors record information asynchronously from one another. The most common of these sensors is the Dynamic Vision System (DVS). A DVS pixel outputs an event $\langle x, y, p, t\rangle$ at the exact timestamp $t$ when its instantaneous log intensity increases or decreases (indicated by $p$) beyond a given threshold \cite{dvs}. These sensors achieve microsecond temporal resolution, high dynamic range, and low power consumption \cite{survey}. However, the output data is large, unwieldy, and difficult to lossy-compress. Furthermore, the sensor does not record the absolute light intensity for non-moving pixels, and researchers often perform multimodal fusion with low-rate framed sensors \cite{survey}. The Dynamic and Active Vision System (DAVIS) sensor combines traditional framed capture and DVS on a single sensor \cite{survey}, but the data streams for frames and events are separate representations. Since DVS events express intensity \textit{change}, to incur loss on one event has a compounding effect on later events for a given pixel. A DVS event stream is an example of a particular type of event video.

The upcoming Aeveon event sensor overcomes the change-based weaknesses of DVS by recording events which directly specify the incident intensity over a dynamic period of time \cite{aeveon}. Effectively, each pixel has its own dynamically tunable shutter speed (exposure time). Pixels which are less interesting to an application can output events infrequently, whereas pixels of high salience can output events at a high rate.

\subsection{Existing Event Video Frameworks}

Due to the difficulty, time, and expense of acquiring large-scale datasets with an event camera, there have been a number of works related to simulating DVS and DAVIS sensors based on framed video inputs \cite{pmlr-v87-rebecq18a,10.3389/fnins.2021.702765,9523069}. Existing frameworks for event camera data focus on generalizing learning-based application interfaces and evaluation mechanisms for particular event cameras; that is, the input event data representation is unchanged, and only the applications are modular \cite{8784777,8460541,7862386}. Furthermore, these systems often quantize the high-rate event information into a temporally redundant framed representation, obviating many advantages of a sparse representation.  

For framed video sources, traditional codecs prove highly effective at compressing temporally redundant data. While some work has explored compressed-domain applications which can carry through this benefit to realize a speed improvement \cite{learning_freq,compressed_recognition,c3d}, the vast majority of video applications operate on the \textit{decompressed} representation \cite{compressed_vision}. Therefore, the compression and application layers are largely divorced in classical video systems, and compression performance does not correlate directly with application speed. This quality is most severe in systems where the video \textit{does} achieve high temporal compression, such as surveillance and high-speed video. An application may incorporate a differencing mechanism to remove temporal redundancy, but such a scheme does not scale with changes to the input frame rate. For example, a doubling of the frame rate would roughly double the computational time that the application spends  isolating temporal changes between frames. In contrast, if the compressed bitrate directly correlates to the \textit{decompressed} bitrate (i.e., the amount of data the application must process), then researchers can develop applications which are more rate-adaptive and predictable.


\subsection{\adder}

While the exact file format of the Aeveon sensor has not yet been publicized, the underlying floating-point intensity outputs and rate adaptation mechanisms bear extreme similarity to my prior work in event-based video representations. I previously introduced the \textbf{A}ddress, \textbf{D}ecimation, \textbf{$\Delta$}t \textbf{E}vent \textbf{R}epresentation (\adder) as an intermediate event-based representation for a variety of video types \cite{freeman_mmsys23}. With my software suite, one can easily transcode video from framed cameras, existing event cameras, and (with little modification) future event cameras such as Aeveon into the single \adder{} representation. By doing so, one can leverage sophisticated rate adaptation schemes, a camera-agnostic application interface, and source-modeled lossy compression \cite{freeman_mmsys23,freeman2023accelerated}.

The default decompressed representation for \adder{} is an event tuple $\langle x,y,c,D,t\rangle$. The spatial coordinates are represented by $x$ and $y$, and the color channel is $c$. The intensity expressed by an event is found by $I = \frac{2^D}{\Delta t}$, where we obtain $\Delta t$ for an event $e_i$ by subtracting the pixel's last event timestamp, $t_{i-1}$, from the current event timestamp, $t_i$. The value of $D$ is determined automatically through a combination of the pixel brightness, the pixel stability (how recently the intensity has changed), and application-level directives (the importance or salience of the pixel). I offer a visualization of \adder{} in \cref{fig:teaser}, highlighting the sparsity of the representation.

\begin{figure}[h]
        \centering
        \includegraphics[width=1.0\linewidth]{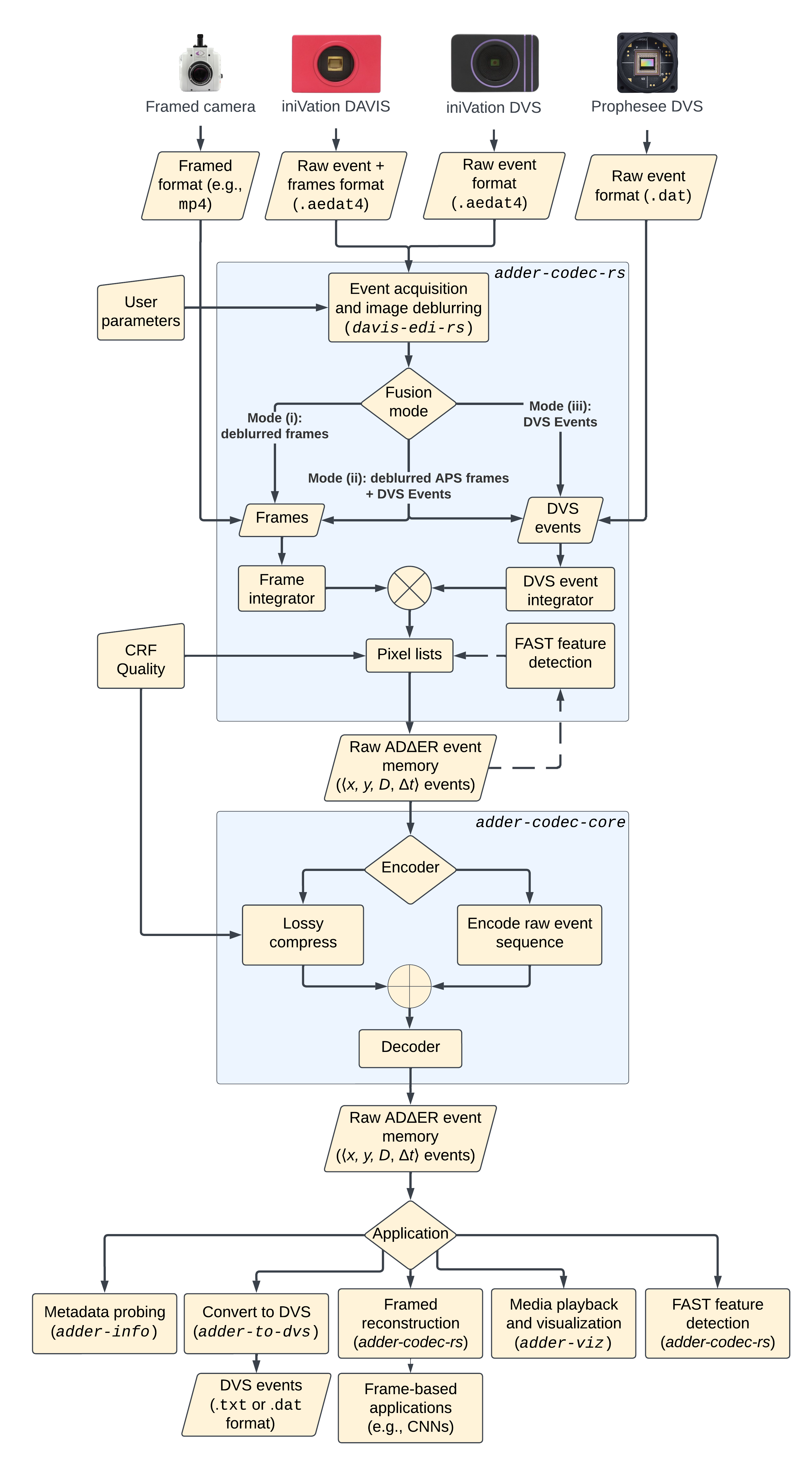}
        \caption{Overview of the \adder{} framework. Italicized names reflect the names of software packages in the Rust Package Registry. This figure is modified and expanded from an earlier version \cite{adder_framework} to include recent additions, such as the CRF quality parameter and Prophesee camera support. }
        \label{fig:software_diagram}
    \end{figure}

\section{Software Architecture}

I designed the \adder{} software to be highly modular. \cref{fig:software_diagram} illustrates the various components and their interdependence. The name of each standalone component is conveyed in italics. I wrote the software in the Rust programming language, and the standalone components are available for download from the Rust Package Registry.


\subsection{Common Codec }\label{sec:core}

The core library (\textit{adder-codec-core} in \cref{fig:software_diagram}) handles the encoding and decoding of \adder{} events, irrespective of their generation. I expose an interface whereby an \adder{} transcoder may instantiate an encoder with a set of options and then simply send its raw events for that encoder to handle. The core can write the raw events directly to a file or stream, or (if the user chooses) queue up event sequences to perform source-modeled lossy compression, as described in \cite{freeman2023accelerated}. Similarly, one may import the core library to act as a decoder for arbitrary \adder{} streams if building a custom application or video player.

To perform lossy compression, the programmer must import the core library with the ``compression'' feature flag enabled. Currently, this feature requires the nightly release channel for Rust, due to unstable features in the subsequent dependency for arithmetic coding. For this reason, the ``compression'' feature is currently disabled by default.

\begin{figure}[h]
        \centering
        \includegraphics[width=1.0\linewidth]{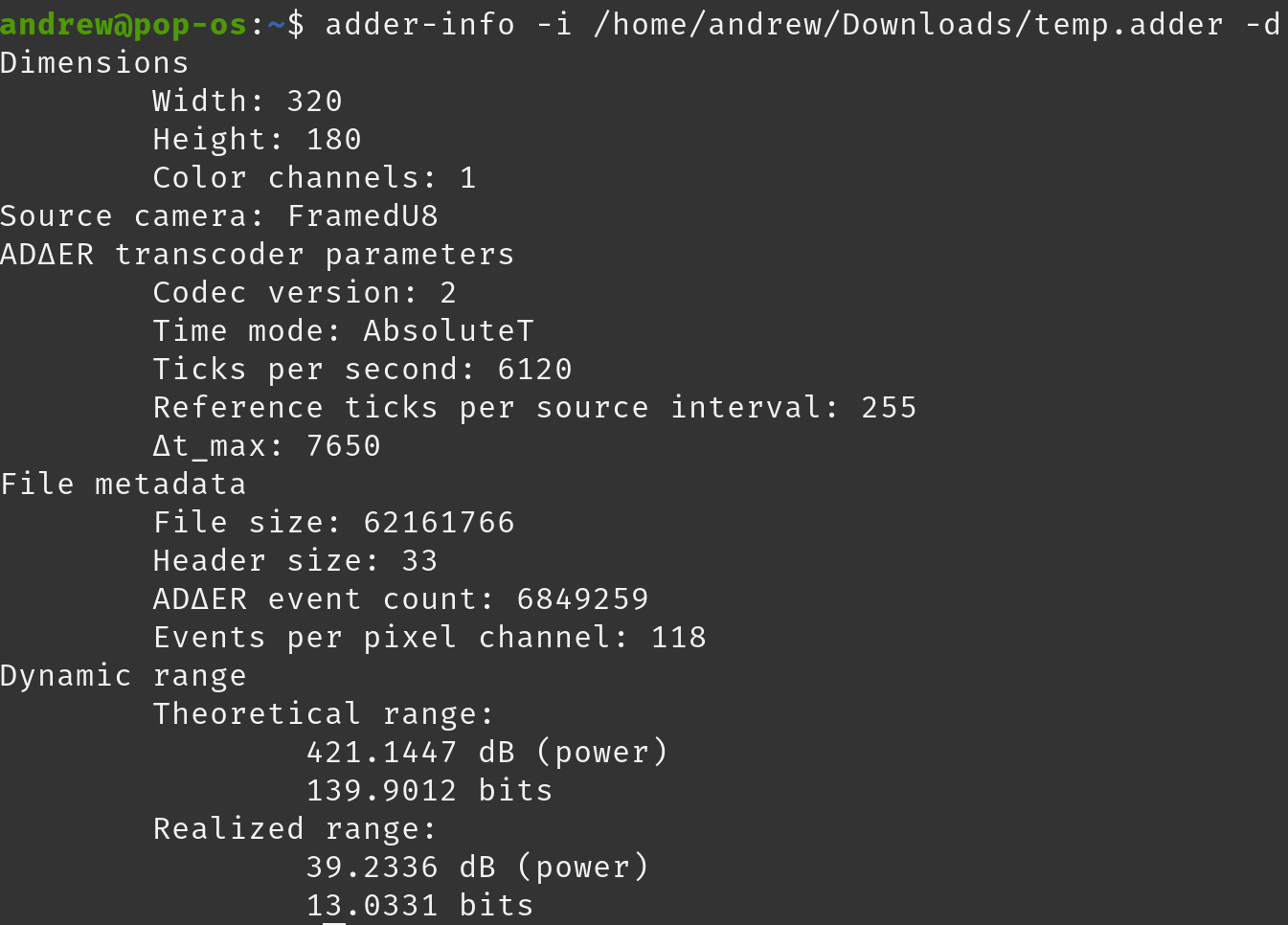}
        \caption{Example output of the \textit{adder-info} utility. The program reports metadata from the file header and calculates the dynamic range of the video.}
        \label{fig:adder_info}
    \end{figure}

\subsection{Metadata Inspection}
The \textit{adder-info} program provides a simple command-line interface to quickly inspect the metadata of an \adder{} file. This program is analogous to the \texttt{ffprobe} utility for framed video \cite{ffmpeg}. It prints information extracted from the file header (encoded by the core library) about the video resolution, time parameters, and video source. Optionally, the program can scan the file to determine the event rate and the dynamic range. In this case, dynamic range refers to the realized precision of the event intensities, given in bits by $\log_2(I_{max}/I_{min})$. Since \adder{} allows stable pixels to average their intensities over time, the precision is often higher than what the source representation allows. Example output is shown in \cref{fig:adder_info}, where the 8-bit framed video source has a reported dynamic range of 13.03 bits in the \adder{} representation. 

\subsection{Middleware}

The programming interfaces for transcoding videos to \adder{}, reconstructing image frames, and running event-based applications are found within the \textit{adder-codec-rs} package.

\subsubsection{Event Generation}

\adder{} currently supports transcoding from frame-based video sources, DVS event sources from camera manufacturers iniVation and Prophesee, and multimodal DVS event \textit{and} framed sources from iniVation DAVIS cameras. The transcoder defines a shared \texttt{Video} interface for generic source video types, including the event encoding mechanism (which calls on \textit{adder-codec-core}, as in \cref{sec:core}), pixel models, and integration functions. A \texttt{Video} has a 3D array (for $x$, $y$, and $c$) of \texttt{EventPixel} structs, which integrate intensity inputs over time to generate \adder{} events. Each \texttt{EventPixel} is independent, determining its optimal $D$ values according to the scheme described in \cite{freeman_mmsys23}.

This scheme uses a linked list to integrate incoming intensities at a range of possible $D$ values. When a node reaches its $2^{D'}$ (for some particular $D'$) intensity threshold, the child of that node is replaced with a new node initialized with decimation $D'$. The parent node stores the generated event in memory and increments its decimation to $D' + 1$ to continue integrating intensities. Then, when the incoming intensity change exceeds the threshold $M$, the \texttt{EventPixel} returns the event stored for each node in the list \cite{freeman_mmsys23}. By design, these events are ordered with monotonically decreasing $D$ and $\Delta t$ values. That is, the first event will have the largest $D$ and implicit $\Delta t$ value, spanning the majority of the integration time.

The multi-node integration process ensures that the full integrated intensity over a long, stable period of time can be precisely represented. However, it can lead to slow performance if recursion is deep, such as when a pixel is very stable and thus has several nodes to integrate. Additionally, the slight variance in intensity precision between a pixel's first event and last event has a negligible effect on reconstruction quality. As such, I introduce a new ``Collapse'' pixel mode as the default integration scheme for \texttt{EventPixel} structs. Under this mode, each pixel integrates \textit{only} a single node, successively incrementing its $D$ when it reaches the integration threshold. When the intensity change exceeds $M$, however, the pixel must account for any time that has elapsed since it generated its candidate event. Therefore, the pixel returns both its candidate event and an ``empty'' event with a reserved $D$ symbol spanning the intervening time. For example, suppose we have an  \texttt{EventPixel} with state $D=9$, $t = 519$, and a running integration of 324 intensity units. The candidate event for the pixel, $\{ D = 8, t = 410\}$, was generated when it reached its last $2^D$ integration, 256. When the incoming intensity changes beyond $M$, this pixel returns the events $\{ D = 8, t = 410\}$ and $\{ D = \texttt{EMPTY}, t = 519\}$. Applications which digest these events will then interpret the latter event as carrying the same average intensity as the first. The collapse mode greatly improves the integration speed, as shown in \cref{tab:perf_framed}.

A particular video source (e.g., framed or DVS) implements the \texttt{Source} trait. Each implementation then defines how to read data a data point from the source representation and convert it into an intensity and timespan. For example, the \texttt{Source} implementation for frame-based video uses an FFmpeg \cite{ffmpeg} backend to decode the next image frame. Then, each 8-bit pixel intensity is integrated in an \texttt{EventPixel} for the same time period (e.g., 255 ticks). In contrast, the multimodal DVS and framed video source must decode packets from a proprietary camera output format, then alternatingly integrate frame-based intensities over a fixed timespan and DVS intensities over variable timespans. 

For event-based video sources, I leverage the \textit{davis-edi-rs} package, which I introduced in \cite{freeman_mmsys23}. This component can ``deblur'' image frames based on their corresponding DVS events, allowing for higher-quality transcodes. Currently, this package depends on OpenCV to process the image frames in log space. One can avoid this dependency by disabling the ``open-cv'' feature flag for the \textit{adder-codec-rs} package, but doing so will remove the ability to transcode from event-based sources. Future work will focus on removing the OpenCV dependency.

\begin{figure*}
  \centering
  \includegraphics[width=1.0\textwidth]{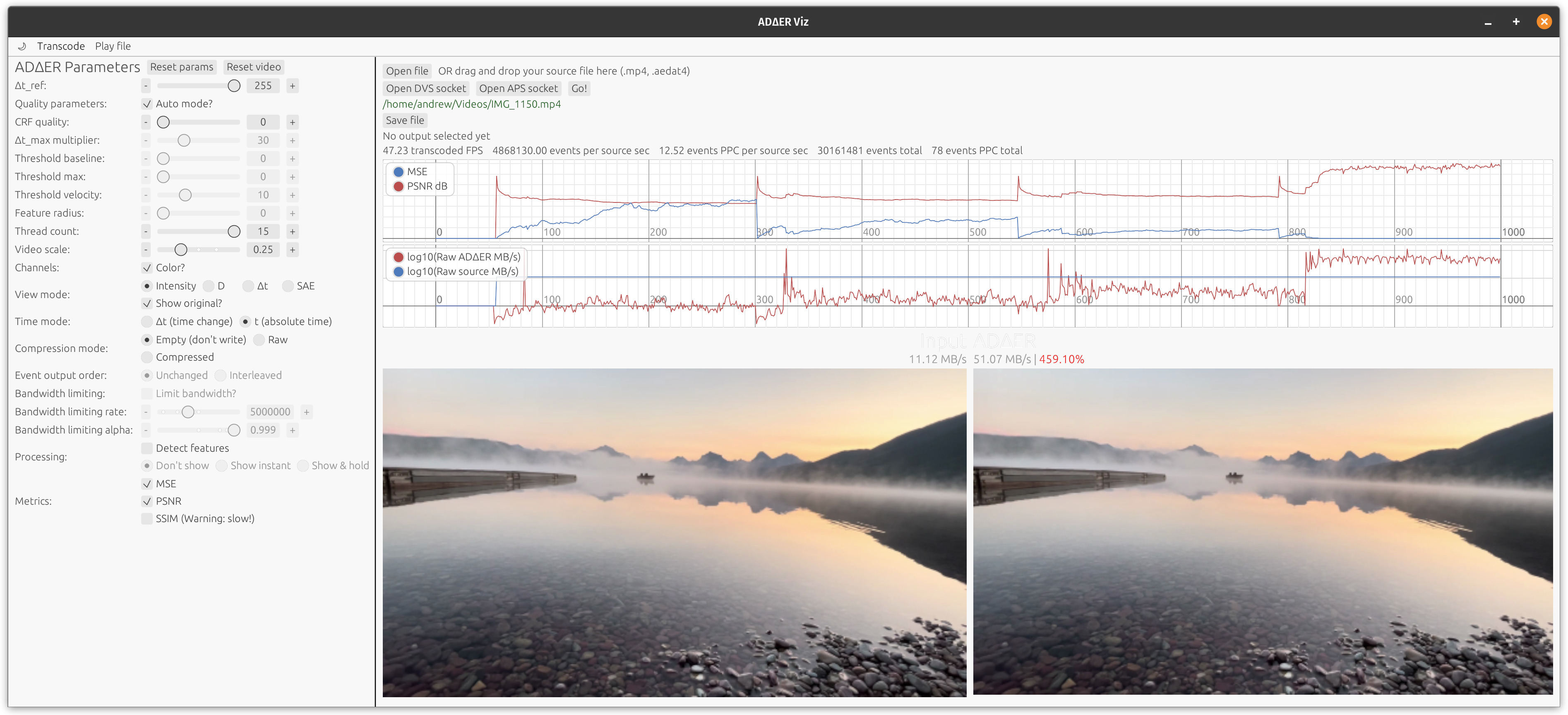}
  \caption{The \textit{adder-viz} transcoder user interface. This short video clip was transcoded to \adder{} at successively higher quality levels. As the CRF level decreases, the quality metrics improve (top graph), but the bitrate increases (bottom graph).}
  \label{fig:transcoder}
\end{figure*}

\subsubsection{Framed Reconstruction}\label{sec:reconstruction}

While \adder{} is asynchronous, display pipelines and most existing vision applications require framed images. As such, this package also provides a reconstructor to generate an image sequence from arbitrary \adder{} events. The user provides an output frame rate which, based on the ticks per second of the \adder{} stream, determines how many ticks each output frame will span. Then, the reconstructor awaits raw events, scaling their intensities to match the timespan of the output frames. When all the pixels in a frame have been initialized with an intensity, the reconstructor outputs that frame. The user may want to visualize only the $D$ or $\Delta t$ components derived from events, rather than the intensities. I support these options by normalizing the desired event components to the range $[0,255]$ (\cref{fig:players}).

\subsubsection{Applications}
Finally, this package contains an implementation of the FAST feature detector as described in \cite{freeman2023accelerated}. I ported and modified the OpenCV FAST detector, which operates on image frames, to instead run on individual pixels. My version receives a pointer to an array which contains the most recent intensity for every pixel, as well as the coordinates of the pixel for which to run the feature test. When the \adder{} event rate is sufficiently low, I found that the event-based version of the algorithm runs upwards of 43\% faster than OpenCV on the VIRAT surveillance dataset \cite{freeman2023accelerated}.

This application (and future applications) can transparently both during \adder{} video playback and while transcoding to \adder{}. In the latter case, one may use the application results to dynamically adjust the pixel sensitivities, allocating available bandwidth towards the pixels of greatest interest. This option is illustrated with the dashed lines in \cref{fig:software_diagram} and described in greater detail in \cite{freeman2023accelerated}. Due to the current low-level integration of the transcoder and applications, incorporating or modifying this application-specific sensitivity adjustment requires changes to the transcoder source code. In the future, I will work to make a modular interface for applications and their effect on transcoder behavior, so that one can develop new applications without delving into the transcoder itself.

The key innovation here is that event-based applications can be, for the first time, agnostic to the imaging modality. My FAST feature detector runs identically on frame-based inputs, DVS inputs from iniVation or Prophesee cameras, and multimodal DVS/APS inputs from iniVation DAVIS cameras. It does not require any tuning or modification for different camera sources. Likewise, \adder{} applications can support any future event-based sensors (such as Aeveon \cite{aeveon}), so long as one implements a simple camera driver and \adder{} transcoder module.


\subsection{Graphical Interface}

Recognizing that the command-line interfaces can be slow and esoteric for new users, I created the \textit{adder-viz} application to enable straightforward explorations of the \adder{} framework. 

\begin{figure*}
     \centering
     \begin{subfigure}[t]{0.33\textwidth}
         \centering
         \includegraphics[width=\textwidth]{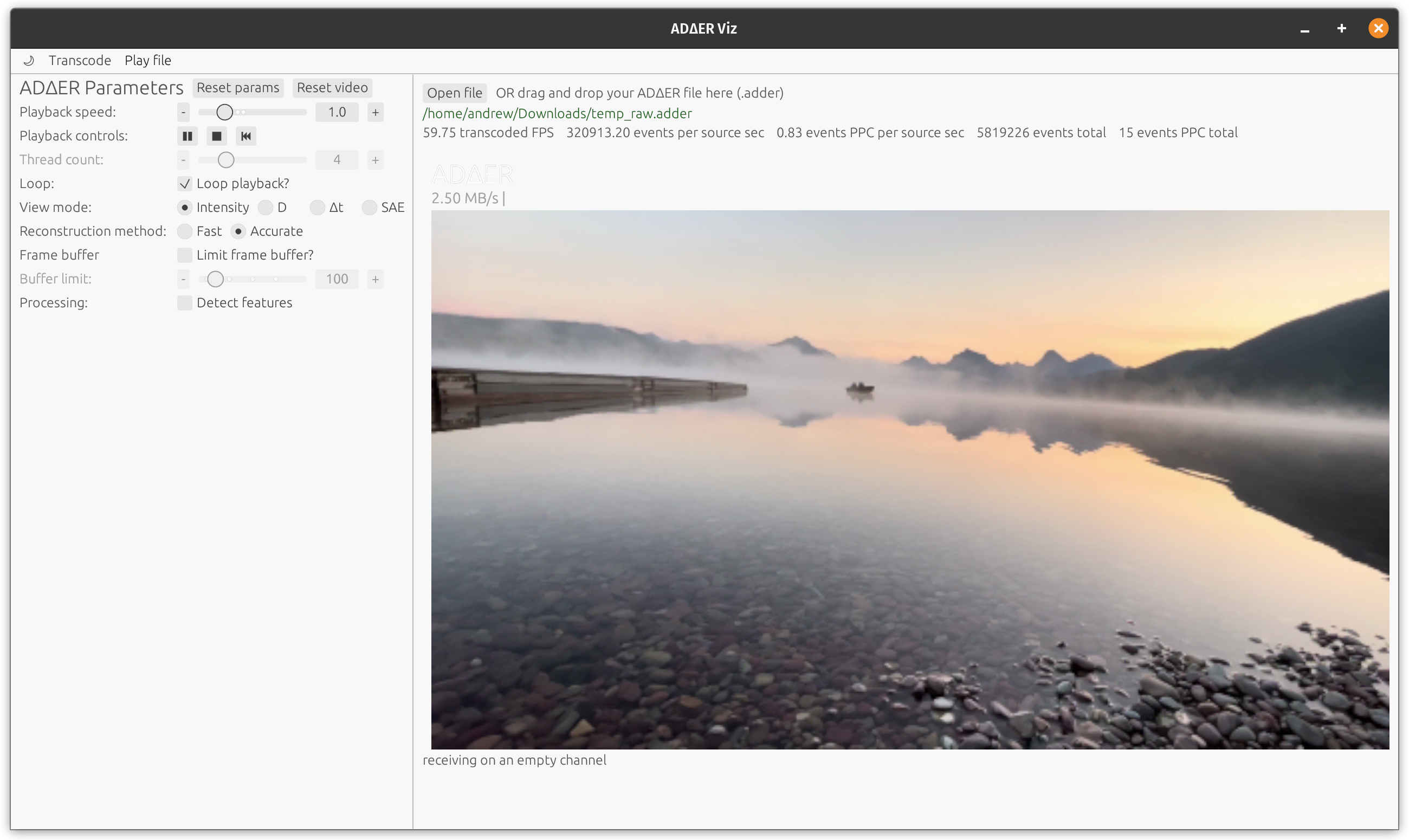}
         \caption{Intensities}
         \label{fig:player_int}
     \end{subfigure}
     \hfill
     \begin{subfigure}[t]{0.33\textwidth}
         \centering
         \includegraphics[width=\textwidth]{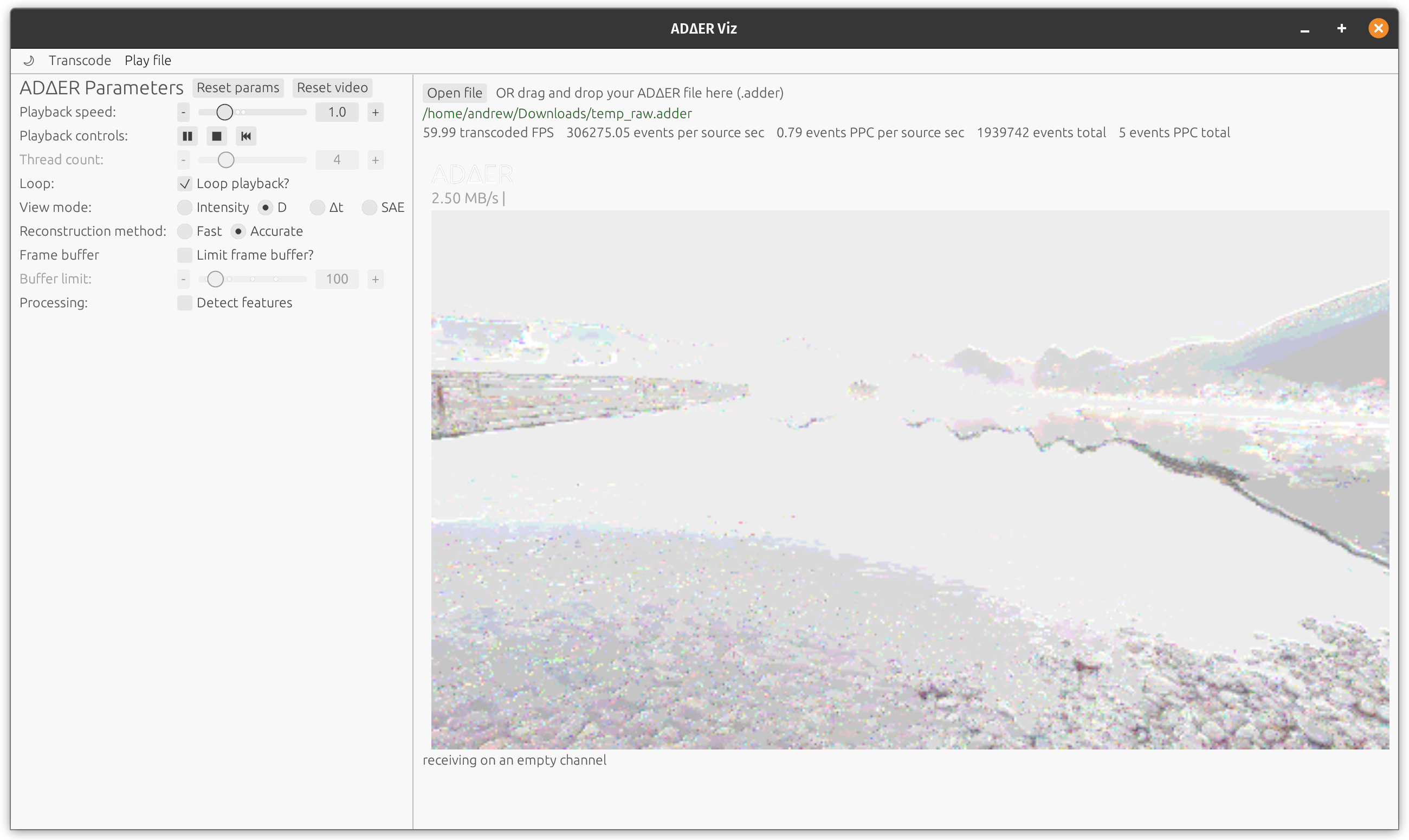}
         \caption{Event $D$ components}
         \label{fig:player_d}
     \end{subfigure}
     \hfill
     \begin{subfigure}[t]{0.33\textwidth}
         \centering
         \includegraphics[width=\textwidth]{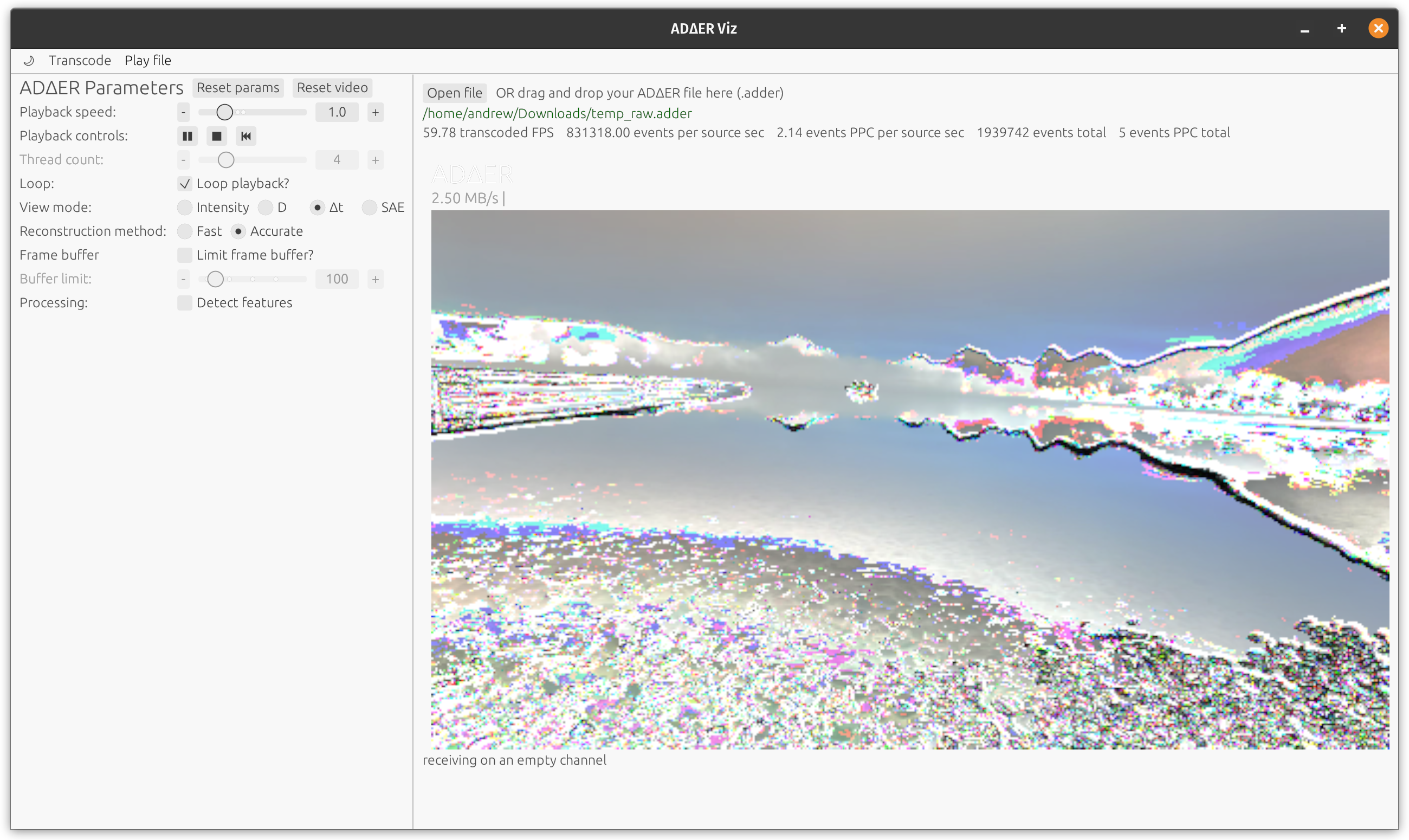}
         \caption{Event $\Delta t$ components}
         \label{fig:player_dt}
     \end{subfigure}
     \caption{The \textit{adder-viz} player interface for \adder{} video, with different visualization modes shown.}
     \label{fig:players}
\end{figure*}

\subsubsection{Transcoder Interface}

The transcoder GUI is shown in \cref{fig:transcoder}. A user can open a framed video file, an AEDAT4 file (from an iniVation-branded DVS camera), or a DAT file (from a Prophesee-branded DVS camera) with a file dialog or by dragging and dropping into the window. Alternatively, a user can open a live connection to a hybrid DVS camera by opening a socket for the DVS events and a socket for the frames. This method requires that the iniVation driver software is running and publishing the data to the respective UNIX sockets. The user can export the \adder{} data by selecting the ``Raw'' or ``Compressed'' output modes and a ``Save file'' destination.

The main panel of the window shows a live view of the transcoded events (the right image in \cref{fig:transcoder}). When the source is a framed video, the application by default displays the input frame on the left. I update the live \adder{} image array with the intensity of a pixel each time it generates a new event. This live update is synchronous with the input, obviating the need for slower framed reconstruction (\cref{sec:reconstruction}).

The left panel shows the various transcoder parameters that a user can adjust. These include settings related to the time representation, pixel sensitivities, resolution, color, and feature-driven rate adaption. Many of these settings can be controlled with a single slider for Constant Rate Factor (CRF) quality, as described in prior work \cite{freeman_mmsys23}. Settings specific to DVS/DAVIS camera sources are made available once an appropriate file or socket connection is established \cite{freeman_mmsys23}. These include options related to event-based deblurring of intensity frames from DAVIS \cite{Pan_EDI}. The user can also enable event-based FAST feature detection and visualize the detected features on the live image.

I provide a number of metrics which are visualized above the display views in \cref{fig:transcoder}. The top plot shows frame-based quality metrics, which the user can enable if the source is a framed video. These include mean squared error (MSE), peak signal-to-noise ratio (PSNR), and structural similarity index measure (SSIM). The bottom plot illustrates the decompressed bitrates of the source and the \adder{} representations. The user can reference these plots to see, in real time, the effect of changing \adder{} transcode parameters on quality and bitrate. For example, \cref{fig:transcoder} shows that our transcoded representation has a lower decompressed bitrate than the source video at many lossy quality levels, but a higher bitrate at the lossless quality level.


\subsubsection{Playback Interface}

\cref{fig:players} shows the video playback interface. One can select a file through a drag-and-drop interaction or a file explorer prompt. The user can pause the video, adjust the playback speed, and visualize the $D$ and $\Delta t$ event components. Furthermore, my event-based FAST feature detection application \cite{freeman_mmsys23} is also available during playback. 

The player supports two playback modes: accurate and fast. The accurate mode leverages the framed reconstruction technique describe in \cref{sec:reconstruction}. This method yields the best visual quality, but may introduce high latency. For example, if some pixel is stable for the duration of a video, it will fire only one event near the beginning of the video encoding. The reconstructor  does not have \textit{a priori } knowledge on whether the pixel has additional events in the future, so it must build a queue of frames for the \textit{entire} video before playback begins. To mitigate the high latency this may cause, the user has an option to limit the size of the frame buffer, such that the reconstructor will assume that a pixel intensity has not changed if its last event sufficiently long ago. In contrast, the fast playback mode simply holds a single image array and updates each pixel intensity when it decodes a new event for that pixel. The player then updates the displayed frame once the change in $t$ represented by any event exceeds a fixed frame interval threshold. Thus, the method is beholden to the temporal event order within the file, which is not guaranteed to be perfectly ordered between different pixels. The lower latency allowed by this method, however, makes it suitable for vision applications, whereas the accurate playback mode is a better suited visualization for human viewers.


\begin{table}
    \centering
    \begin{tabular}{cc|cc|cc}
        & & \multicolumn{2}{c|}{Raw events} & \multicolumn{2}{c}{Lossy compression} \\
          &
         Resolution & Grayscale & Color & Grayscale & Color \\
        \hline \parbox[t]{2mm}{\multirow{4}{*}{\rotatebox[origin=c]{90}{Normal}}} 
        & 480×270   & 209.0 & 109.7  & 153.9   & 71.1 \\
        & 960×540   & 71.6  & 33.8  &  43.2  & 17.1 \\
        & 1440×810  & 32.9  & 15.7  & 19.6   & 7.6 \\
        & 1920×1080 & 22.7  & 9.5  &  11.5  & 4.1 \\
    \hline \parbox[t]{2mm}{\multirow{4}{*}{\rotatebox[origin=c]{90}{Collapse}}} 
        & 480×270   & 262.3 & 161.6 & 176.6 & 88.9 \\
        & 960×540   & 91.8  & 48.4  & 51.3 & 20.5 \\
        & 1440×810  & 43.2  & 22.2  & 23.0 & 9.0 \\
        & 1920×1080 & 31.1  & 13.8  & 13.5 & 4.7 \\
        
    \end{tabular}
    \caption{Frames per second at which a representative framed video can be transcoded to \adder{}. Resolution, color depth, and lossy compression are varied. A CRF value of 3 (the default) was used for these experiments.}
    \label{tab:perf_framed}
\end{table}

\section{Performance and Future Work}

I gathered general speed measurements on a machine with a Ryzen 5800x CPU with 8 cores and 16 threads. As shown in \cref{tab:perf_framed}, the \adder{} transcoder achieves fast performance for low-resolution framed inputs, but slows substantially at high definition. At 540p resolution and higher, the time required to transcoding a color video is about double that of the grayscale version.

I use a memory-safe parallelization scheme for matrix integrations (transcoding image frames) and framed reconstruction, whereby pixel arrays are divided into groups of spatial rows. Currently, however, most other processes are serial. According to performance profiling results, roughly $75\%$ of transcoder execution time is spent on pixel integrations. Each software pixel is a struct that is dynamically sized according to the length of its event queue. The transcoder logic will likely benefit from a GPU implementation, though this may require a static maximum queue length unless the ``Collapse'' mode is used. This may prove most advantageous on GPUs with direct memory access, so that high-rate event sequences do not have to move through the CPU cache during encoding or decoding.

Currently, the lossy compression scheme is single-threaded, and it carries a computational overhead over raw event encoding (\cref{tab:perf_framed}). Within the \textit{adder-viz} GUI, this manifests as a notable pause each time an application data unit (ADU) of events undergoes lossy compression. This latency may be mitigated if compression were delegated to its own background thread and if a second ADU were constructed while one is being compressed. Furthermore, compression speed can likely be improved by dividing the events into horizontal spatial regions for parallel processing.

While this tool suite provides an end-to-end system for event-based video, there is ample room to extend it with additional features and performance improvements. I will also work to create a generic application interface and simplify the application integration process for new researchers.  Finally, I will work to incorporate \adder{} events as inputs for novel spiking neural network vision applications, which can take advantage of sparse representations.

\section{Conclusion}

This open-source release and user guide for the \adder{} framework provides researchers with straightforward tools to experiment with forward-looking event video. As the sensor community pushes for ever higher resolution, dynamic range, sample rate, and event-based representations, \adder{} unlocks novel approaches to rate adaptation, compression, and applications.

\begin{acks}
This work is partially supported by a grant from the United States Department of Defense.
\end{acks}

\bibliographystyle{ACM-Reference-Format}
\bibliography{BIBLIOGRAPHY}










\end{document}